\documentclass[aps,prc,preprint]{revtex4-1}

\usepackage{units}

\usepackage{url}

\usepackage{amssymb} 
\usepackage{amsmath}

\usepackage{graphicx}

\usepackage{color}

\def\be{\begin{equation}}
\def\ee{\end{equation}}
\def\bea{\begin{eqnarray}}
\def\eea{\end{eqnarray}}
\def\beas{\begin{eqnarray*}}
\def\eeas{\end{eqnarray*}}

\newcommand{\zbb}{2\mbox{$\nu\beta\beta$-decay}}

\newcommand{\pdhz}{\mbox{$^{110}$Pd}~}

\newcommand{\nel}{\mbox{$\nu_e$}}
\newcommand{\bpbp}{\ensuremath{\beta^+\beta^+}~}
\newcommand{\bec}{\ensuremath{\beta^+/{\rm EC}}}
\newcommand{\ecec}{\ensuremath{{\rm EC/EC}}~}
\newcommand{\ema}{\ensuremath{\langle m_{\nu_e} \rangle}~}

\begin{document}
\newcommand{\nuc}[2]{$^{#2}\rm #1$}

\newcommand{\bb}[1]{$\rm #1\nu \beta \beta$}
\newcommand{\bbm}[1]{$\rm #1\nu \beta^- \beta^-$}
\newcommand{\bbp}[1]{$\rm #1\nu \beta^+ \beta^+$}
\newcommand{\bbe}[1]{$\rm #1\nu \rm ECEC$}
\newcommand{\bbep}[1]{$\rm #1\nu \rm EC \beta^+$}

\newcommand{\pic}[5]{
       \begin{figure}[ht]
       \begin{center}
       \includegraphics[width=#2\textwidth, keepaspectratio, #3]{#1}
       \end{center}
       \caption{#5}
       \label{#4}
       \end{figure}
}

\newcommand{\apic}[5]{
       \begin{figure}[H]
       \begin{center}
       \includegraphics[width=#2\textwidth, keepaspectratio, #3]{#1}
       \end{center}
       \caption{#5}
       \label{#4}
       \end{figure}
}

\newcommand{\sapic}[5]{
       \begin{figure}[P]
       \begin{center}
       \includegraphics[width=#2\textwidth, keepaspectratio, #3]{#1}
       \end{center}
       \caption{#5}
       \label{#4}
       \end{figure}
}

\newcommand{\picwrap}[9]{
       \begin{wrapfigure}{#5}{#6}
       \vspace{#7}
       \begin{center}
       \includegraphics[width=#2\textwidth, keepaspectratio, #3]{#1}
       \end{center}
       \caption{#9}
       \label{#4}
       \vspace{#8}
       \end{wrapfigure}
}

\newcommand{\baseT}[2]{\mbox{$#1\cdot10^{#2}$}}
\newcommand{\baseTsolo}[1]{$10^{#1}$}
\newcommand{\THL}{$T_{\nicefrac{1}{2}}$}

\newcommand{\UBI}{$\rm cts/(kg \cdot yr \cdot keV)$}

\newcommand{\Uflux}{$\rm m^{-2} s^{-1}$}
\newcommand{\Ucpd}{$\rm cts/(kg \cdot d)$}
\newcommand{\Uexpo}{$\rm kg \cdot d$}

\newcommand{\Qbb}{$\rm Q_{\beta\beta}\ $}

\newcommand{\validate}{\textcolor{blue}{\textit{(validate!!!)}}}

\newcommand{\improve}{\textcolor{blue}{\textit{(improve!!!)}}}

\newcommand{\missing}{\textcolor{red}{\textbf{...!!!...} }}

\newcommand{\quanta}{\textcolor{red}{\textit{(quantitativ?) }}}

\newcommand{\misscite}{\textcolor{red}{[citation!!!]}}

\newcommand{\missref}{\textcolor{red}{[reference!!!]}\ }

\newcommand{\PC}{$N_{\rm peak}$}
\newcommand{\BIC}{$N_{\rm BI}$}
\newcommand{\PAPR}{$R_{\rm p/>p}$}

\newcommand{\PCR}{$R_{\rm peak}$}


\newcommand{\gline}{$\gamma$-line}
\newcommand{\glines}{$\gamma$-lines}


\newcommand{\tab}{Tab.~}
\newcommand{\eq}{Eq.~}
\newcommand{\fig}{Fig.~}
\renewcommand{\sec}{Sec.~}
\newcommand{\chap}{Chap.~}

 \newcommand{\fn}{\iffalse \fi} 
 \newcommand{\tx}{\iffalse \fi} 
 \newcommand{\txe}{\iffalse \fi} 
 \newcommand{\sr}{\iffalse \fi} 


\title{New Half-life Limits on Double Beta Decays of \nuc{Pd}{110} and \nuc{Pd}{102} into Excited States}



\author{Bj\"{o}rn Lehnert}
\email{bjoern.lehnert@tu-dresden.de}
\affiliation{Institut f\"{u}r Kern- und Teilchenphysik\\ Technische Universit\"{a}t Dresden, Germany}

\author{Kai Zuber}
\email{Zuber@physik.tu-dresden.de}
\affiliation{Institut f\"{u}r Kern- und Teilchenphysik\\ Technische Universit\"{a}t Dresden, Germany}

\author{Erica Andreotti}%
\email{Erica.ANDREOTTI@ec.europa.eu}
\affiliation{Institute for Reference Materials and Measurements, Retieseweg 111, B-2440 Geel, Belgium}

\author{Mikael Hult}
\email{Mikael.HULT@ec.europa.eu}
\affiliation{Institute for Reference Materials and Measurements, Retieseweg 111, B-2440 Geel, Belgium}



\date{\today}

\begin{abstract}
A search for excited state transitions of double beta decay in \nuc{Pd}{110} and double electron capture in \nuc{Pd}{102} has been performed in the HADES underground laboratory with two low background HPGe detectors in sandwich configuration. After an exposure of \unit[35.92]{kg$\cdot$d}, no signal was found. The frequentist spectral analysis resulted in lower half-life limits of \unit[\baseT{1.98}{20}]{yr} and \unit[\baseT{1.72}{20}]{yr} (\unit[95]{\%} CL) for the first $0^+$ and $2^+$ excited state in \nuc{Pd}{110} respectively. This is an improvement by more than a factor of 3 with respect to previous measurements. In \nuc{Pd}{102}, the lower half-life limit could be improved to \unit[\baseT{5.95}{18}]{yr} (\unit[95]{\%} CL) for the first $0^+$ excited state. Furthermore, first experimental lower half-live limits are found for all possible excited states in the \nuc{Pd}{110} and \nuc{Pd}{102} systems.\\[1pc]

Structured Abstract:\\
\noindent
{\bf Background:} Excited state transitions in double beta decays are a powerful tool to validate and tune calculations of nuclear matrix elements. \\

\noindent
{\bf Purpose:} The experimental lower half-life limits for double beta decays of \nuc{Pd}{110} and \nuc{Pd}{102} into the excited $2^+_1$ and $0^+_1$ are improved. Furthermore, first limits of transitions into the $2^+_2$, $0^+_2$ and $2^+_3$ states are published for \nuc{Pd}{110} as well as a first limit for the $2^+_2$ state transition in \nuc{Pd}{102}.\\

\noindent
{\bf Methods:} The Pd sample was measured with two HPGe detector in sandwich configuration in the HADES underground laboratory during \unit[44.77]{d} of life time. The analysis is performed with the frequentist Fieldman Cousisns method.\\

\noindent
{\bf Results:} Lower half-life limits of \unit[\baseT{1.98}{20}]{yr} and \unit[\baseT{1.72}{20}]{yr} (\unit[95]{\%} CL) have been found for the first $0^+$ and $2^+$ excited state in \nuc{Pd}{110} respectively. This is an improvement by more than a factor of 3 with respect to previous measurements. In \nuc{Pd}{102}, the lower half-life limit could be improved to \unit[\baseT{5.95}{18}]{yr} (\unit[95]{\%} CL) for first $0^+$ excited state. Furthermore, first experimental lower half-live limits are found for all possible excited states in the \nuc{Pd}{110} and \nuc{Pd}{102} systems\\

\noindent
{\bf Conclusions:} Previous half-life limits have been improved and experimental results were obtained for all theoretical calculations of palladium double beta decays into excited states. \\

\end{abstract}


\maketitle

\section{Introduction}

The field of neutrino physics has made groundbreaking progress in recent years. Oscillation experiments studying neutrinos coming from the sun \cite{Abe2011,SNOCollaboration2011,Alimonti2009}, the atmosphere \cite{Hatakeyama1998}, nuclear reactors \cite{Kim2012a,An2012,Abe2008,Lasserre2006} and in accelerator beams \cite{Adamson2008,Abe2011a,Agafonova2012} have found compelling evidence for flavor oscillation in the lepton sector. This changes the long believed assumption of neutrinos being mass-less particles. The implication from neutrino oscillation, that at least two neutrino mass eigenstates have a non-zero rest mass, does not allow fixing an absolute mass-scale and leaves two mass hierarchy scenarios open. Advanced oscillation experiments try to identify the hierarchy scenario by using oscillation effects in the earth or in dense stellar matter \cite{Mathews2012} but are not able to determine the absolute mass scale. This has to be done via beta-decay \cite{Wolf2010,Monfardini2006a}, cosmology \cite{Jarosik2011} or neutrinoless double-beta decay: 

\begin{equation}
\label{eq:1}
(Z,A) \rightarrow (Z+2,A) + 2 e^-  \quad (0\nu\beta\beta) \, , 
\end{equation}

which would have a high sensitivity to determine the mass scale of neutrinos. This process violates total lepton number by two units and thus is not allowed in the Standard Model. Furthermore, it is the gold plated process to distinguish whether neutrinos are Majorana or  Dirac particles.

For the \bb{0} process to exist, it is necessary to match the helicities of the intermediate neutrino states  which is easiest done by introducing a neutrino mass. This mass is connected to the experimentally observable half-life via

\begin{equation}
  \label{eq:1}
 \left(T_{1/2}^{0 \nu}\right)^{-1} = G^{0 \nu}(Q, Z) \cdot \left| M_{GT}^{0\nu} - M_{F}^{0\nu} \right|^2 \cdot \left(\frac{\ema}{m_e}\right)^2 \, ,
\end{equation}

where \ema is the effective Majorana neutrino mass given by the coherent sum over the virtual electron neutrino mass eigenstates $\ema = \left| \sum_{i}U_{ei}^2m_i\right|$ with $U_{ei}$ as the lepton flavor mixing matrix, $G^{0 \nu}(Q, Z) $ is a phase space factor and $M_{GT}^{0\nu} - M_{F}^{0\nu}$ describes the nuclear transition matrix element. The experimental signature is the emission of two electrons with a sum energy corresponding to the Q-value of the nuclear transition. A potential evidence has been claimed for the \bb{0} mode of  $^{76}$Ge with $T_{1/2}^{0 \nu} = \unit[2.23^{+0.44}_{-0.31} \times 10^{25}]{yr}$ at \unit[90]{\%} CL \cite{Klapdor2001,Klapdor2006}. \\

Experimentally observed in 11 nuclides \cite{TRETYAK2002,Ackerman2011c} is the SM process of neutrino accompanied double beta decay:

\begin{equation} 
(Z,A) \rightarrow (Z+2,A) + 2 e^- + 2 \nu_e \quad (2\nu\beta\beta) \, ,
\end{equation}

which is expected with half-lives around \unit[$10^{20}$]{yr} depending on the Q-value. For recent reviews see \cite{Avignone2008}. An alternative process is the double positron decay in combination with electron capture (EC). 
Three different decay modes can be considered:

\begin{align}
(Z,A) \rightarrow& (Z-2,A) + 2 e^+ + (2 \nel)  &\mbox{(\bpbp{})}\\
e^- + (Z,A) \rightarrow& (Z-2,A) + e^+ + (2 \nel) & \mbox{(\bec{})}\\
2 e^- + (Z,A) \rightarrow& (Z-2,A) + (2 \nel)  &\mbox{(\ecec{})}
\end{align}

Decay modes containing a positron have a reduced Q-value as each generated positron accounts for a reduction 
of 2 $m_ec^2$ and thus can only occur in nuclides with sufficient energy difference to the daughter nuclide. The full energy is only available in the \ecec\ mode and makes it the most probable one; however, it is also the most difficult to detect, only producing X-rays instead of \unit[511]{keV} gammas.\\

Another set of searches focuses on excited state transitions in double beta decays. This is experimentally interesting because the event topology is enhanced by de-excitation gammas which are accessible with gamma ray spectroscopy. However, without information about the other final state particles, this technique cannot distinguish between the \bb{0} and the \bb{2} mode; hence, the deduced half-lives are valid for both. The investigation of \bb{2} modes into excited states provides information on nuclear structure which is valuable for matrix element calculations. The understanding and tuning of parameters in matrix elements for \bb{2} modes is imperative for translating the measured half-life of \bb{0} experiments into a Majorana neutrino mass (\eq \ref{eq:1}). So far only transitions to the first excited $0^+$ state have been observed in \nuc{Mo}{100} \cite{Barabash1995} and \nuc{Nd}{150} \cite{Barabash2004}. The searches described in this paper are searches for excited state transitions in palladium with gamma ray spectroscopy. \\

\section{Double Beta Decays in Palladium}

The element under study is palladium with the isotopes of interest \nuc{Pd}{110} and \nuc{Pd}{102}. Among the 35 isotopes expected to undergo $\beta^-\beta^-$ decay, \nuc{Pd}{110} has the second highes natural abundance with \unit[11.72]{\%}. Recently, the Q-value was remeasured to \unit[2017.85(64)]{keV} \cite{Fink2012a} and places \nuc{Pd}{110} among the 11 $\beta^-\beta^-$ isotopes with a Q-value larger than \unit[2000]{keV}. Two measurements of \nuc{Pd}{110} have been performed in the past in 1954 \cite{Winter1952} and more recently in 2011 \cite{Lehnert2011}. The latter measurement was the first to investigate excited states in palladium and is the direct predecessor of this search. There exist many theoretical calculations for transitions into the ground state and into the $2^+_1$, $0^+_1$, $2^+_2$, $0^+_2$ and $2^+_3$ excited states to which the experimental limits can be compared. The existing experimental and theoretical half-life limits ares summarized in \tab \ref{tab:table2}. This paper has the aim to provide experimental information about every decay mode into excited states in \nuc{Pd}{110} and \nuc{Pd}{102} that has been investigated theoretically. Furthermore, \pdhz is an excellent candidate to probe the single-state dominance hypothesis for \zbb, i.e.\ that only the lowest lying intermediate 1$^+$-state will contribute to the nuclear transition matrix element describing its \zbb.\\

The second isotope \nuc{Pd}{102} has a Q-value of \unit[1172]{keV}, a natural abundance of \unit[1.02]{\%} and is able to decay via \ecec and \bec. The only experimental half-life limit was quoted in \cite{Lehnert2011} and no theoretical calculation have been published up to date. A summary can be found in \tab \ref{tab:table3}.\\

\begin{table}[]
\caption{\label{tab:table2}  Experimental and theoretical half-life limits for various $\beta\beta$ decay modes in \nuc{Pd}{110}. The columns show from left to right the theoretical model the quoted half-life, the reference and the year of publication. Abbreviations denote:
PHFM - Projected Hartree-Fock-Bogoliubov,
SSDH - Single-State-Dominance Hypothesis,
SRPA - Second quasi Random Phase Approximation,
OEM - Operator Expansion Method,
QRPA - Quasi Random Phase Approximation,
SSD - Single State Dominance and
pnQRPA - proton-neutron quasiparticle random-phase approximation.
}
\begin{ruledtabular}
\scriptsize
\begin{tabular}{cccc}
 Exp / Th model & Lower Limit \THL\ [yr] & Reference & Year of Publication\\
\hline
\multicolumn{ 3}{l}{\nuc{Pd}{110} Ground State Transition}\\
\hline
Exp  & \baseT{1}{17} (\unit[68]{\%} CL) &  \cite{Winter1952}  & 1952\\
PHFM &  \baseT{1.41}{20} and \baseT{3.44}{20} \footnotemark[1]  &  \cite{Chandra2005} & 2005\\
SSDH &  \baseT{1.75}{20}  &  \cite{Semenov2000}  & 2000\\
SSDH &  \baseT{1.2-1.8}{20} \footnotemark[2]   &  \cite{Civitarese1998} & 1998 \\
SRPA &  \baseT{1.6}{20}    &  \cite{Stoica1994} & 1994\\
OEM &  \baseT{1.24}{21}    &  \cite{Hirsch1994} & 1994 \\
QRPA &  \baseT{1.16}{19}   &  \cite{Staudt1990} & 1990\\
SSD &   \baseT{1.2}{20}   & \cite{Domin2005} & 2005\\
pnQRPA &   \baseT{1.1}{20} and \baseT{0.91}{20} \footnotemark[3]   & \cite{Suhonen2011} & 2011 \\
\hline
\multicolumn{ 3}{l}{\nuc{Pd}{110} $2^+_1$ Excited State Transition @ \unit[657.76]{keV}}\\
\hline
Exp &  \baseT{4.40}{19} (\unit[95]{\%} CL)  & \cite{Lehnert2011} & 2011\\
SSD &  \baseT{4.4}{25}   &  \cite{Domin2005}  & 2005\\
SRPA &   \baseT{8.37}{25}    & \cite{Stoica1994} & 1994 \\
pnQRPA &   \baseT{1.48}{25}   &  \cite{Raduta2007}  & 2007\\
pnQRPA &   \baseT{0.62}{25} and \baseT{1.3}{25} \footnotemark[3]   &  \cite{Suhonen2011} & 2011 \\
\hline
\multicolumn{ 3}{l}{\nuc{Pd}{110} $0^+_1$ Excited State Transition @ \unit[1473.12]{keV}}\\
\hline
Exp &  \baseT{5.89}{19} (\unit[95]{\%} CL) & \cite{Lehnert2011} & 2011\\
SSD &  \baseT{2.4}{26}    &  \cite{Domin2005} & 2005\\
pnQRPA &  \baseT{4.2}{23} and \baseT{9.1}{23} \footnotemark[3]  &  \cite{Suhonen2011} &2011 \\
\hline
\multicolumn{ 3}{l}{\nuc{Pd}{110} $2^+_2$ Excited State Transition @ \unit[1475.80]{keV}}\\
\hline
SSD  &  \baseT{3.8}{31}  &  \cite{Domin2005} & 2005\\
pnQRPA &   \baseT{11}{30} and \baseT{7.4}{30} \footnotemark[3] &  \cite{Suhonen2011} & 2011\\
\hline
\multicolumn{ 3}{l}{\nuc{Pd}{110} $0^+_2$ Excited State Transition @ \unit[1731.33]{keV}}\\
\hline
SSD  &   \baseT{5.3}{29}  &  \cite{Domin2005} & 2005\\
\hline
\multicolumn{ 3}{l}{\nuc{Pd}{110} $2^+_3$ Excited State Transition @ \unit[1783.48]{keV}}\\
\hline
SSD  &  \baseT{1.3}{35}  &  \cite{Domin2005}  & 2005\\
 
\end{tabular}
\end{ruledtabular}
\footnotetext[1]{$g_A$ = 1.25 and 1.0 respectively}
\footnotetext[2]{Different experimental input for calculations}
\footnotetext[3]{For Woods-Saxon Potential and adjusted base respectively (see \cite{Suhonen2011} for details)}
\end{table}

\begin{table}[]
\caption{\label{tab:table3} Experimental half-life limits for various ECEC and $\beta^+{\rm EC}$ decay modes in \nuc{Pd}{102}.\\
}
\begin{ruledtabular}
\footnotesize
\begin{tabular}{ccccc}
 Exp / Th model & Lower Limit \THL\ [yr]  & Reference & Year of Publication\\
\hline
\multicolumn{ 3}{l}{\nuc{Pd}{102} $2^+_1$ Excited State Transition @ \unit[475.10]{keV}}\\
\hline
Exp &   \baseT{2.68}{18} (\unit[95]{\%} CL) &  \cite{Lehnert2011} &2011 \\
\hline
\multicolumn{ 3}{l}{\nuc{Pd}{102} $0^+_1$ Excited State Transition @ \unit[943.69]{keV}}\\
\hline
Exp & \baseT{7.64}{18} (\unit[95]{\%} CL)  &  \cite{Lehnert2011}  &2011\\

\end{tabular}
\end{ruledtabular}

\end{table}

\begin{figure}[h]
 \includegraphics[width=0.99\textwidth]{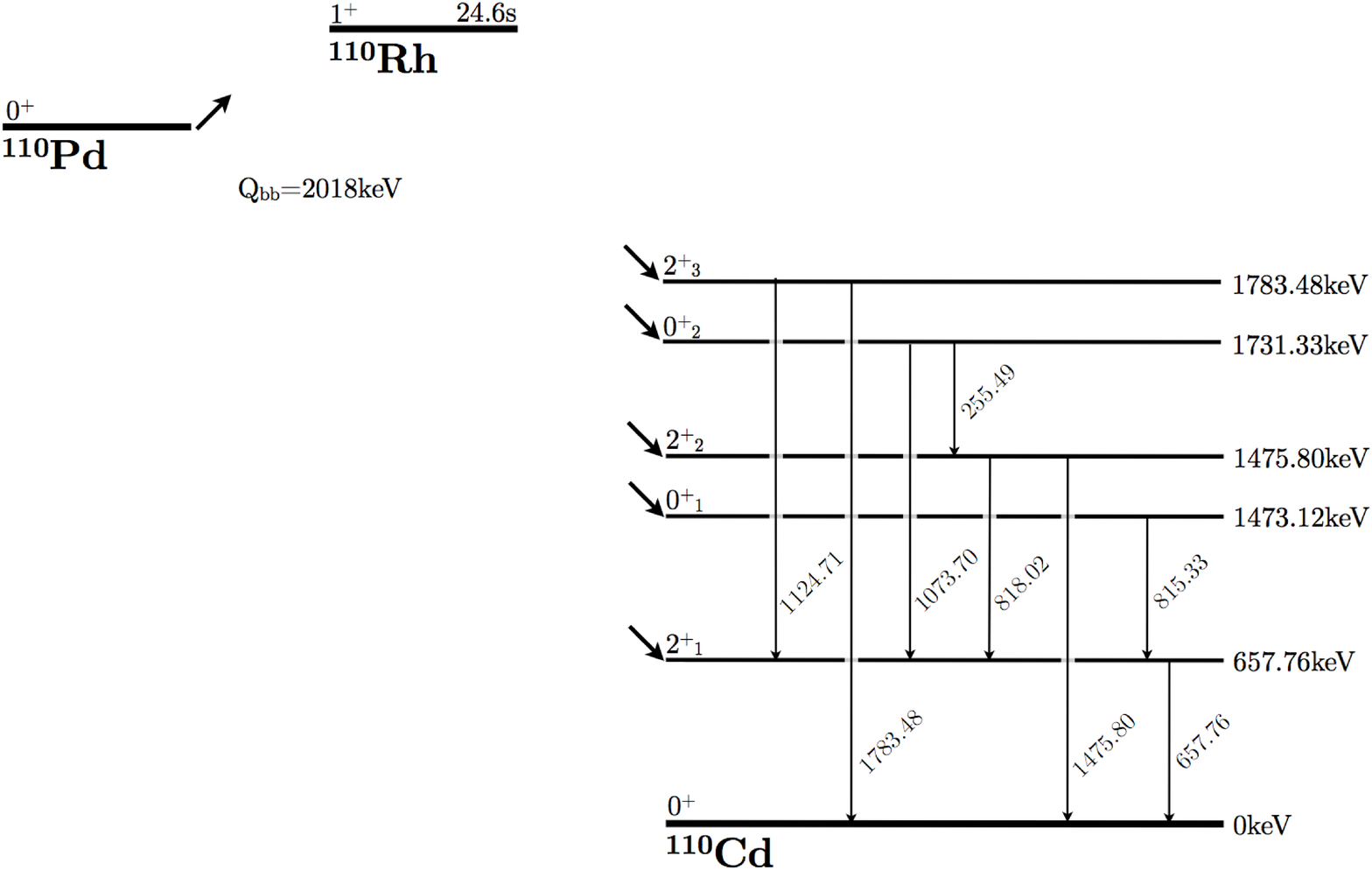}%
 \caption{ \label{fig:levelSchemePd110} \nuc{Pd}{110} level scheme of investigated decay modes. Nuclear data from \cite{ENSDF2012}.}
 \end{figure}

\begin{figure}[h]
 \includegraphics[width=0.99\textwidth]{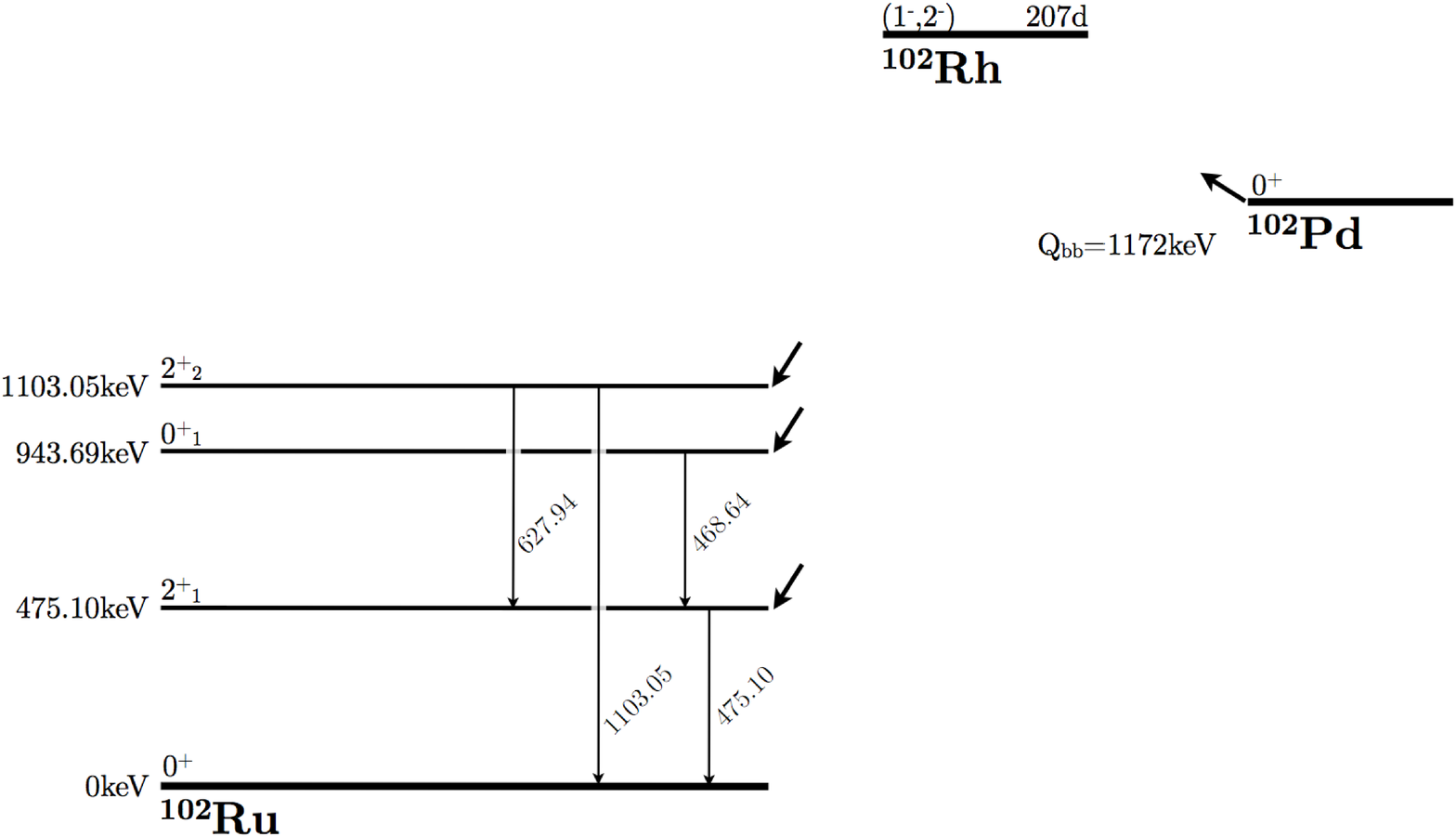}%
 \caption{ \label{fig:levelSchemePd102} \nuc{Pd}{102} level scheme of investigated decay modes. Nuclear data from \cite{ENSDF2012}.}
 \end{figure}

The search is based on gamma spectroscopy, hence only gamma lines are considered in the event topology. Each excited state transition is followed by an unique set of decay branches and gamma cascades which are illustrated in \fig \ref{fig:levelSchemePd110} and \ref{fig:levelSchemePd102}. 

\section{Experiment}

\subsection{Measuring Setup}

The measurements were performed in the High Activity Disposal Experimental Site, HADES, underground laboratory on the premises of the Belgian Nuclear Research Centre SCK$\cdot$CEN in Mol. The underground laboratory is located at a depth of \unit[225]{m} inside a Boom clay formation and has a flat overburden that amounts to  roughly \unit[500]{m w.e.} \cite{HADES}.\\

\begin{figure}[h]
 \includegraphics[width=0.9\textwidth]{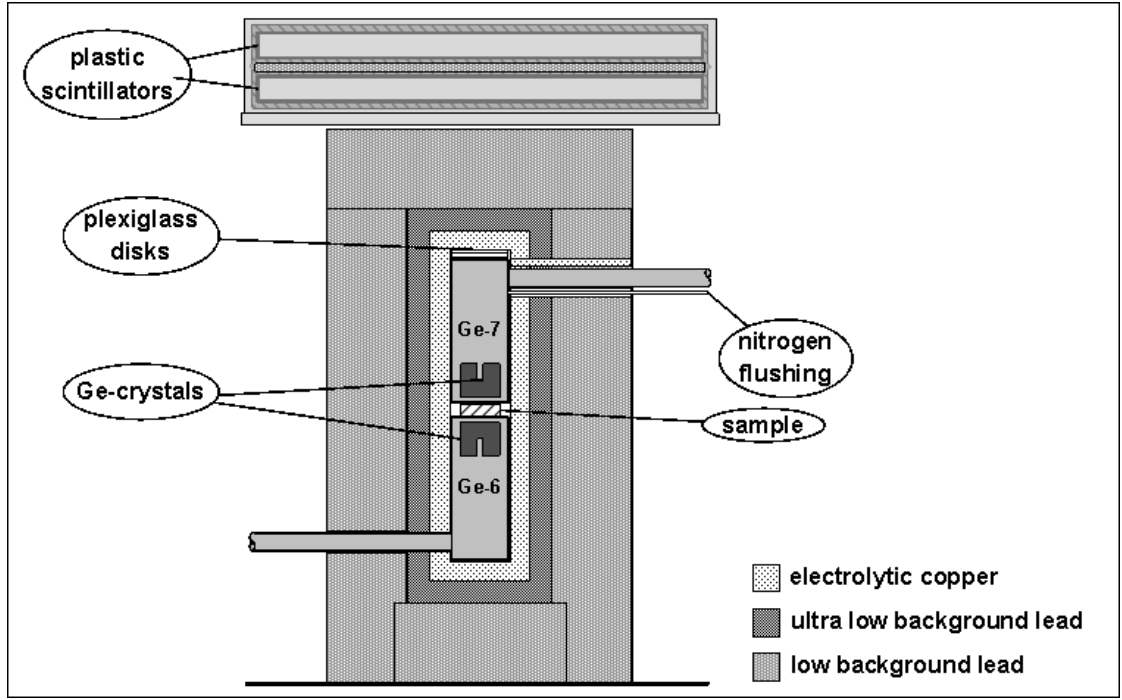}%
 \caption{ \label{fig:detector} IRMM germanium detector sandwich setup.} 
 \end{figure}

The detector setup consists of two HPGe detectors in a sandwich configuration with integrated muon veto panels on the top \cite{Wieslander2009a}. It is shown in \fig \ref{fig:detector}. The sample is placed between the top (Ge-7) and bottom (Ge-6) detector. The distance between the two can be adjusted in order to maximize the solid angle acceptance and the detection efficiency. Ge-6 is a p-type HPGe detector with \unit[80]{\%} efficiency and \unit[0.9]{mm} dead layer in a \unit[102]{mm} cryostat with Cu endcap whereas Ge-7 is an extended range p-type HPGe detector with  \unit[90]{\%} efficiency and \unit[0.3]{$\mu$m} dead layer in an Al cryostat. The characteristics of Ge7 makes it suitable for the detection of low energetic X-rays while the configuration of Ge6 has the advantage of  reducing the background and X-ray coincidences. The shielding consists of an outer layer of \unit[14.5]{cm} \unit[20]{Bq/kg} (\nuc{Pb}{210}) lead, an intermediate layer of  \unit[4.0]{cm} \unit[2.4]{Bq/kg} (\nuc{Pb}{210}) low activity lead and an inner layer of  \unit[3.5]{cm} electrolytic copper with less than \unit[15]{$\mu$Bq/kg} \nuc{Co}{60} and less than \unit[20]{$\mu$Bq/kg} \nuc{Th}{228} \cite{Wieslander2009a}.\\

The data acquisition is two-fold. The main DAQ is the IRMM self-fabricated DAQ2000 multi-parameter system that is recording events in list mode from the two HPGe detectors as well as from the two muon panels and enables coincidence analysis in the ROOT framework \cite{Antcheva2009}. Additionally, a standard GENIE DAQ system was used in histogram mode for each HPGe detector redundantly. The DAQ2000 was only operative for a reduced measuring time with limited sample exposure; thus the analysis in this work is performed with the GENIE DAQ and without muon veto or detector coincidence. \\

The total background rate in the germanium detectors was previously measured with \unit[992]{cts/d} in an energy range of \unit[40..2400]{keV} of which \unit[124]{cts/d} were identified as muon events \cite{Wieslander2009a}.

\subsection{Palladium Sample}

\begin{figure}[h]
 \includegraphics[width=0.95\textwidth]{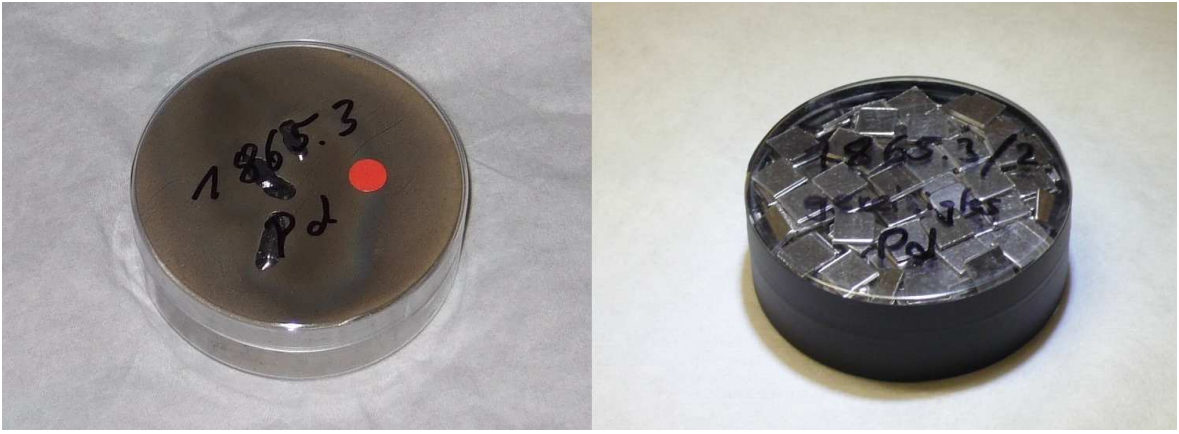}
 \caption{ \label{fig:PdSample} Palladium sample before (left) and after (right) purification inside the measuring container.}
 \end{figure}

\unit[802.35]{g} of irregular shaped \unit[$1$]{mm} x \unit[1]{cm$^2$} plates of palladium were placed inside a measuring container of \unit[70]{mm} diameter and \unit[50]{mm} height. The plates are piled inside the container as dense as possible and an effective density of \unit[10.2]{g/cm$^3$} was calculated. The palladium is approximated with a homogeneous distribution and the effective density in the simulations for determining the detection efficiency. Recently, the sample was purified by C. HAFNER GmbH + Co. KG in 2010 to a certified purity of \unit[$>99.95$]{\%} which lowered the continuous background in the peak regions by approximately \unit[20]{\%} \cite{Lehnert2011}. In order to avoid radioisotopes produced by cosmic ray spallation, the palladium was kept underground and exposed only \unit[18]{d} during purification in 2010 and \unit[3]{d} for transport in fall 2011. A picture of the palladium sample before and after purification is shown in \fig \ref{fig:PdSample}.

\section{Analysis}

\subsection{Stability Check}
The histogram data of the GENIE DAQ is separated into individual runs of roughly \unit[24]{h} for each of the two detectors Ge6 and Ge7. This enables the stability check of the DAQ system over the extended period of measurement and the removal of individual runs.\\

The stability was checked by plotting the count rate and peak centroid of the background peaks at \unit[609]{keV} and \unit[2614]{keV} as a function of time (spectrum number). Although counting statistics prevented detailed analysis of the short term stability, it was clear that there was no measurable energy drift during the measurement period. In addition, quality controls with a point source containing \nuc{Co}{60}, \nuc{Cs}{137} and \nuc{Am}{241} were carried out before, after and once in-between the measurement. The average \nuc{Rn}{222} activity concentration in the laboratory during the data taking was measured to \unit[9.5]{Bq/m$^3$} and at no point higher than \unit[25]{Bq/m$^3$}. No correlation to the background count rate was observed mainly due to effective Rn-removal by minimizing empty space inside the shield and flushing with N$_{2}$. The count rate remained stable over the 32 runs with one exception; in the first two runs an increase in total counts and counts from \nuc{Tl}{208} was observed, however not from \nuc{Bi}{214}. This behavior was cross checked with additional visible \glines\ from \unit[583.19]{keV} \nuc{Tl}{208} and \unit[239.63]{keV} \nuc{Pb}{212}, and \unit[1764.49]{keV} \nuc{Bi}{214} representing the \nuc{Th}{232} and \nuc{U}{238} decay chains respectively. The increased count rate of the \nuc{Th}{232} chain in the first days is only seen in the lower Ge6 with a Cu endcap and not in the upper Ge7 with an Al endcap. One possible explanation is that the short half-life of \nuc{Rn}{220} (\unit[55.6]{s}) implies that it is not flushed out by boil-off nitrogen and that its daughters preferentially stick to the copper surface of the lower detector (Ge6) than to the aluminum surface of the upper detector (Ge7). \nuc{Rn}{222} with longer half-life (\unit[3.8]{d}) will be flushed out of the shield before a significant amount of daughters are produced.

Consequently, the first two runs are removed from the analysis resulting in a total of 30 runs with \unit[44.77]{d} of good data out of 32 runs with \unit[46.49]{d} total data which translates into a total exposure of \unit[35.92]{kg$\cdot$d}.

\subsection{Data Processing}

\begin{figure}[h]
 \includegraphics[width=0.95\textwidth]{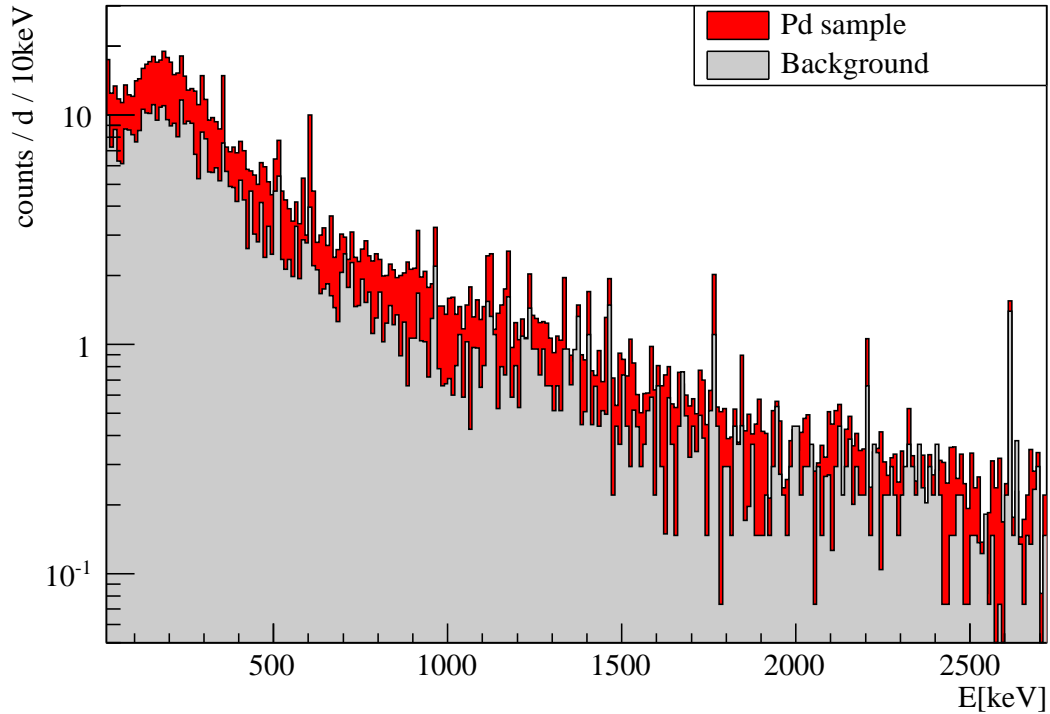}
 \caption{ \label{fig:SumSpec} Sum spectrum of Ge6 and Ge7 for selected data set (\unit[44.77]{d}) in red and background spectrum (\unit[13.62]{d}) in gray. The spectra are shown in common \unit[10]{keV} bins between \unit[20 and 2720]{keV}.}
 \end{figure}

For the final analysis a single energy spectrum is used for which all individual runs are summed: In a first step, all selected runs of one detector are combined with the same energy calibration. In a second step, the single detector spectra are rebinned into a common binning of \unit[0.5]{keV/bin} and a common energy range from \unit[20..2720]{keV}. Additionally, the Ge6 spectrum is scaled to the life-time of the Ge7 spectrum which becomes the common life-time of the sum spectrum. The commonly binned and scaled spectra are added. These steps result in a non-integer sum spectrum that denotes the count per bin in the life-time of Ge7. The difference in life-time between the two detectors is less than \unit[0.5]{\%} for the selected runs. The combined spectrum of Ge6 and Ge7 together with a background spectrum of \unit[13.62]{d} is shown in \fig \ref{fig:SumSpec}.

\subsection{Background Investigation}

The radiopurity of the palladium sample was assessed during the $\gamma$-spectrometry measurement performed in HADES. In the measured spectrum, the major \glines\ emitted by natural radionuclides belonging to the \nuc{U}{238}, \nuc{Th}{232} chains and \nuc{K}{40} as well as the \nuc{Co}{60} lines are visible. The palladium spectrum is compared to the background spectrum measured without a sample. The background peak count rate is subtracted from that of the palladium sample. The result, if positive, is then used for the evaluation of the activity due to the impurities in the sample. In case of a negative result, a decision threshold is calculated according to \cite{ISO_11929_2010}. The results are reported in \tab \ref{tab:bgResults}. Decision thresholds are also calculated for the following radionuclides: \nuc{Rh}{102} (\THL = \unit[207.3]{d}), \nuc{Rh}{102m} (\THL = \unit[3.742]{yr}) and \nuc{Ag}{110m} (\THL = \unit[249.76]{d}). The reason is the possible interference with the search for \nuc{Pd}{110} and \nuc{Pd}{102} isotopes, because of the emission of \glines\ from the same excited daughter states. No presence of these radionuclides is found, as reported in \tab \ref{tab:bgIntermResults}.

\begin{table}[]
\caption{\label{tab:bgResults} Massic activities (in \unit{mBq/kg}) of radioimpurities detected in the Pd sample.}
\begin{ruledtabular}
\footnotesize
\begin{tabular}{lcccc}

\multicolumn{1}{c}{} & \multicolumn{1}{c}{} & \multicolumn{1}{c}{} & \multicolumn{1}{c}{Decision Threshold} & \multicolumn{1}{c}{Weighted Mean} \\ 
\multicolumn{1}{c}{Nuclide} & \multicolumn{1}{c}{E} & \multicolumn{1}{c}{Massic Activity} & \multicolumn{1}{c}{ ($\alpha=\unit[95]{\%}$)} & \multicolumn{1}{c}{Massic Activity} \\ 
 & \multicolumn{1}{c}{[keV]} & \multicolumn{1}{c}{[mBq/kg]} & \multicolumn{1}{c}{[mBq/kg]} & \multicolumn{1}{c}{[mBq/kg]} \\ 
\hline
\nuc{Pb}{214} & 295.22 & 1.9 $\pm$ 1.0 & 1.4 & 1.4 $\pm$ 0.4\\ 
\multicolumn{ 1}{l}{} & 351.93 & 1.3 $\pm$ 0.5 & 0.6 & \multicolumn{ 1}{r}{} \\ 
\hline
\nuc{Bi}{214} & 609.32 & 1.9 $\pm$ 0.4 & 0.4 & 1.9 $\pm$ 0.4 \\ 
\multicolumn{ 1}{l}{} & 1120.29 & 2.0$\pm$ 0.8 & 0.9 & \multicolumn{ 1}{r}{} \\ 
\multicolumn{ 1}{l}{} & 1238.11 & --- & 2.2 & \multicolumn{ 1}{r}{} \\ 
\multicolumn{ 1}{l}{} & 1377.67 & --- & 2.7 &  \\ 
\multicolumn{ 1}{l}{} & 1764.54 & --- & 3.2 &  \\ 
\hline
\nuc{Pb}{210} & 46.54 & --- & 414.3 &  \\ 
\hline
\nuc{Ac}{228} & 911.20 & --- & 0.5 &  \\ 
\multicolumn{ 1}{l}{} & 968.97 & --- & 0.9 &  \\ 
\hline
\nuc{Pb}{212} & 238.63 & --- & 0.7 &  \\ 
\hline
\nuc{Tl}{208} & 583.19 & --- & 0.6 &  \\ 
\multicolumn{ 1}{l}{} & 2614.51 & --- & 0.3 &  \\ 
\hline
\nuc{K}{40} & 1460.82 & --- & 1.0 &  \\ 
\hline
\nuc{Cs}{137} & 661.66 & --- & 0.2 &  \\ 
\hline
\nuc{Co}{60} & 1173.23 & --- & 0.2 &  \\ 
\multicolumn{ 1}{l}{} & 1332.49 & --- & 0.1 &  \\ 
\end{tabular}

\end{ruledtabular}

\end{table}

\begin{table}[]
\caption{\label{tab:bgIntermResults} Decision thresholds for direct $\gamma$-background of DBD intermediate nuclei.\\
}
\begin{ruledtabular}
\footnotesize
\begin{tabular}{lrc}

\multicolumn{1}{c}{}   & & \multicolumn{1}{c}{Decision Threshold}  \\ 
\multicolumn{1}{c}{Nuclide} & \multicolumn{1}{c}{E} & \multicolumn{1}{c}{ ($\alpha=\unit[95]{\%}$)}  \\ 
 & \multicolumn{1}{c}{[keV]} & \multicolumn{1}{c}{[mBq/kg]}  \\ 
\hline
\nuc{Ag}{110m} & 1384.30 &   0.5\\ 
\multicolumn{ 1}{l}{} & 1505.04 &  1.0 \\ 

\nuc{Rh}{102} & 475.05 &   0.4 \\ 
\nuc{Rh}{102m} & 631.28 &  0.3\\ 
\multicolumn{ 1}{l}{} & 697.49 & 0.3  \\ 
\end{tabular}

\end{ruledtabular}

\end{table}

\subsection{Peak Finding}

The analysis is a peak search on the detector sum spectrum and either retrieves the number of counts in a respective peak or states an upper limit of counts according to a level of confidence.

In an experiment with non negligible background, the background can fluctuate upwards or downwards. The sensitivity of an experiment is then defined as a resulting signal which originates from a \unit[1]{$\sigma$} upward fluctuation of the background. This can be calculated before performing the experiment if the background is known. In case of the observation of a downward fluctuation, which formally results in negative signal counts, the signal is usually set to zero and the sensitivity is quoted as an upper limit of the counts. On the other hand, when using classical uncertainties on the observed downward fluctuated counts, it is possible that in some cases even the upper limit its negative. In these cases it results in a poor coverage of the quoted confidence level at best and in an unphysical negative result at worst. Additionally there is a discontinuity in the coverage when crossing from a two-sided confidence interval definition to a one-sided one. 

All these problems are addressed by the method of Feldman and Cousins in their paper \cite{Feldman1998}. They use a Neyman construction of a confident belt and an ordering principle based on likelihood ratios. The advantages are a physical yield in all background situations i.e.\ positive upper count limit and avoiding discontinuities in the coverage while crossing the statistical interpretation from a non-observation to an observation, i.e.\ from a one-sided to a two-sided confidence interval. The confidence intervals of the Feldman Cousins method are believed to have a better coverage for small numbers than gaussian ones \cite{Feldman1998}.\\ 

No prominent peak structures are observed in the signal region and the results of the analysis are upper limits only for the peak counts. In order to obtain a numerical value, all bins within a peak are combined into a single analysis bin that covers at least the FWHM of the peak. The real signal fraction coverage is calculated as the gaussian peak area in the analysis bin and depends on the actual binning of the spectrum. The peak background is fitted with a constant functions defined \unit[$\pm 30$]{keV} around the peak energy excluding a window of \unit[$\pm 5$]{keV} around the peak. In case of prominent background peaks in the side bands, they are included in the background function as gaussians. This was done for the background peaks at \unit[609.31]{keV}, \unit[1120.29]{keV} and \unit[1764.49]{keV}  from \nuc{Bi}{214}, at \unit[238.63]{keV} from \nuc{Pb}{212}, at \unit[1461.83]{keV} from \nuc{K}{40} and at \unit[661.66]{keV} from \nuc{Cs}{137}. In case of the \unit[661.66]{keV} peak which is close to the \unit[657.76]{keV} peak in the \nuc{Pd}{110} system, the flat background function was defined closer than \unit[$\pm 5$]{keV} into the signal region to improve the background estimation.\\

The observed counts in the analysis bin are compared to the expected background with the ROOT class TFeldmanCousins, which returns the lower and upper bound of the signal confidence interval according to a specified confidence level that is set to \unit[95]{\%} in this work. All investigated peaks show a lower count limit of zero; this is in agreement with a non-observation of the peak. To account for the incomplete coverage of the peak area by the analysis bin, the upper count limit is divided by the fraction of coverage and thus adjusts the upper count limit in a conservative way.\\

An illustration of the technique is shown in \fig \ref{fig:657keV} to \ref{fig:468keV} with the energy spectrum in the solid black line, the background function in the dashed blue line, the peak fraction marked as the solid red area and the gaussian signal peak as it appears with the Feldman Cousins upper limit as the red solid line.

The Feldman and Cousins limits are cross checked with the ISO standard methods \cite{ISO_11929_2010} and agree better than within a factor of two with each other; this can be explained by the different treatment of statistical background fluctuations.

\subsection{Monte Carlo Simulation for \gline\ Efficiencies}


The full energy peak efficiencies are determined using Monte Carlo simulations with the EGS4 software. The models of the detectors were first determined from manufacturer data and using information from radiography. Thereafter the dead layer thicknesses were adjusted in the model to agree with measured FEP efficiencies from point sources within \unit[3]{\%}. The final model has been validated using volume sources of similar size as the Pd source in this study. Each decay branch is simulated separately with information from \cite{ENSDF2012} and the calculations involved all the cascading $\gamma$-rays of each branch so that the resulting FEP efficiency is inherently corrected for the coincidence summing effect. X-ray coincidences and the angular correlations are neglected in the simulations and it was assumed that the activity was homogeneously distributed in the whole volume of the sample.

\section{Results}

\begin{figure}[h]
 \includegraphics[width=0.9\textwidth]{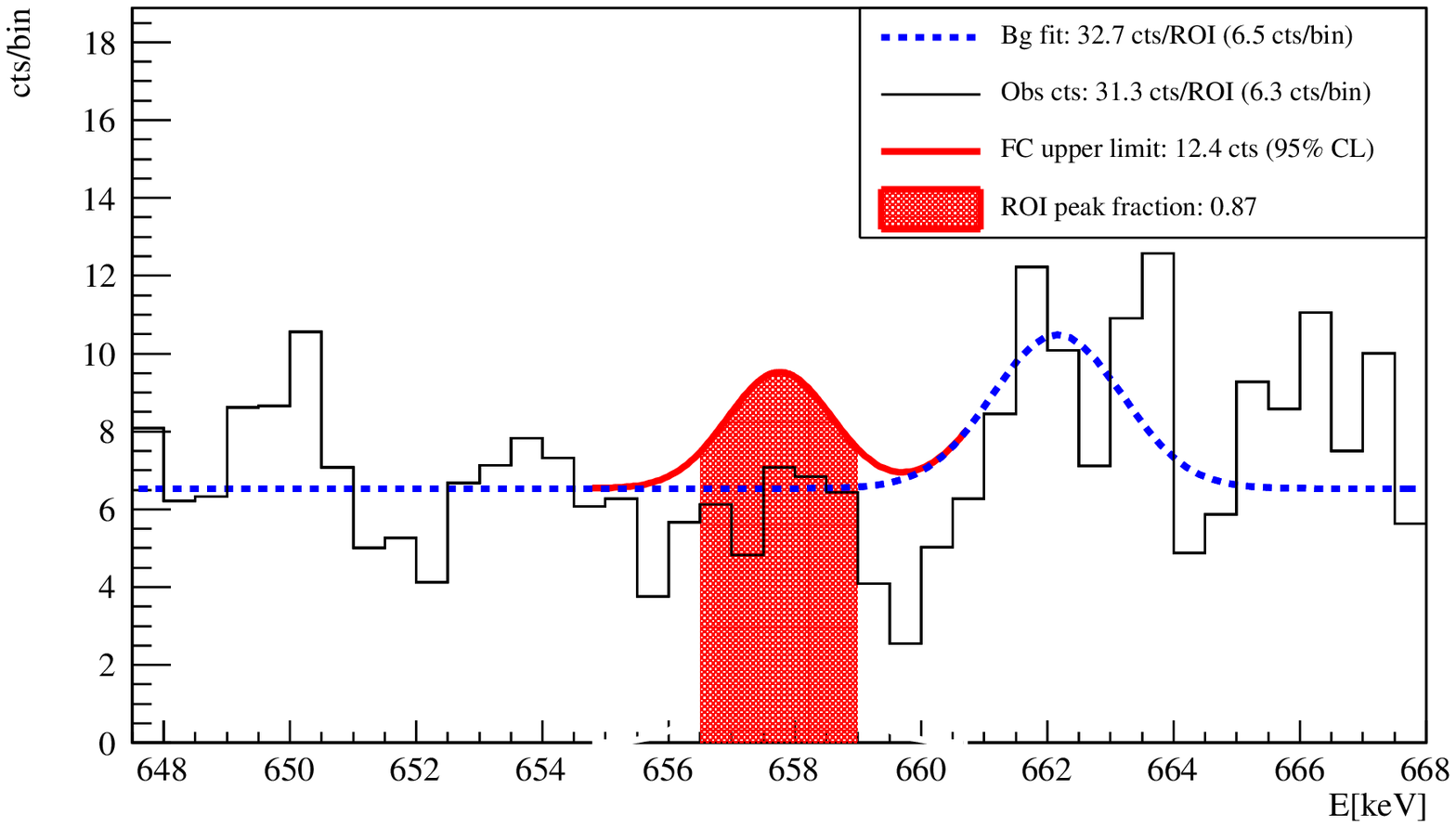}
 \caption{ \label{fig:657keV} The peak region around the  \unit[657.76]{keV} \gline\ in \nuc{Pd}{110} and the \unit[661.66]{keV} \nuc{Cs}{137} \gline.}
 \end{figure}

\begin{figure}[h]
 \includegraphics[width=0.9\textwidth]{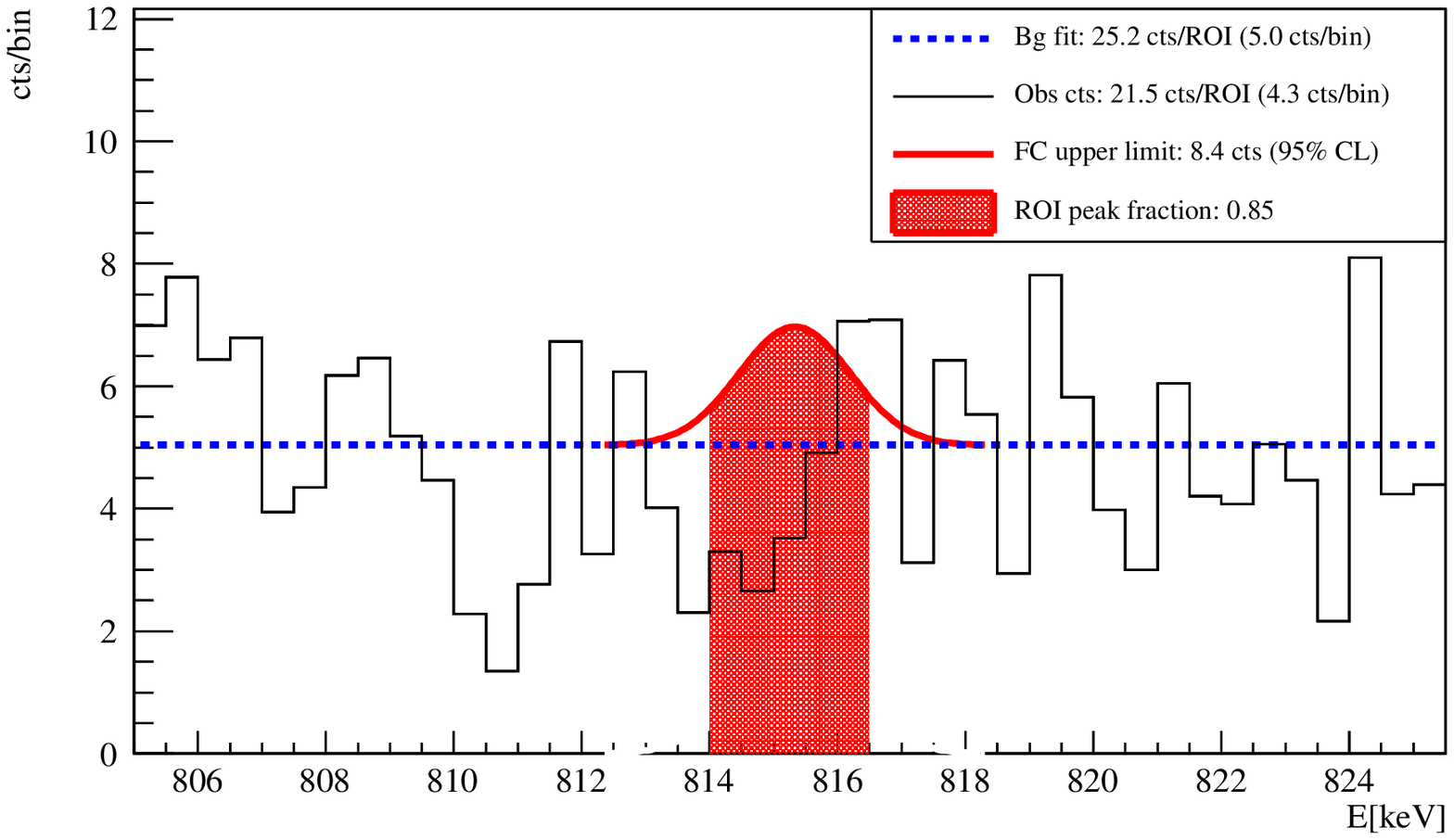}
 \caption{ \label{fig:815keV} The peak region around the \unit[815.33]{keV} \gline\ from \nuc{Pd}{110} $0_1^+$.}
 \end{figure}

\begin{figure}[h]
 \includegraphics[width=0.9\textwidth]{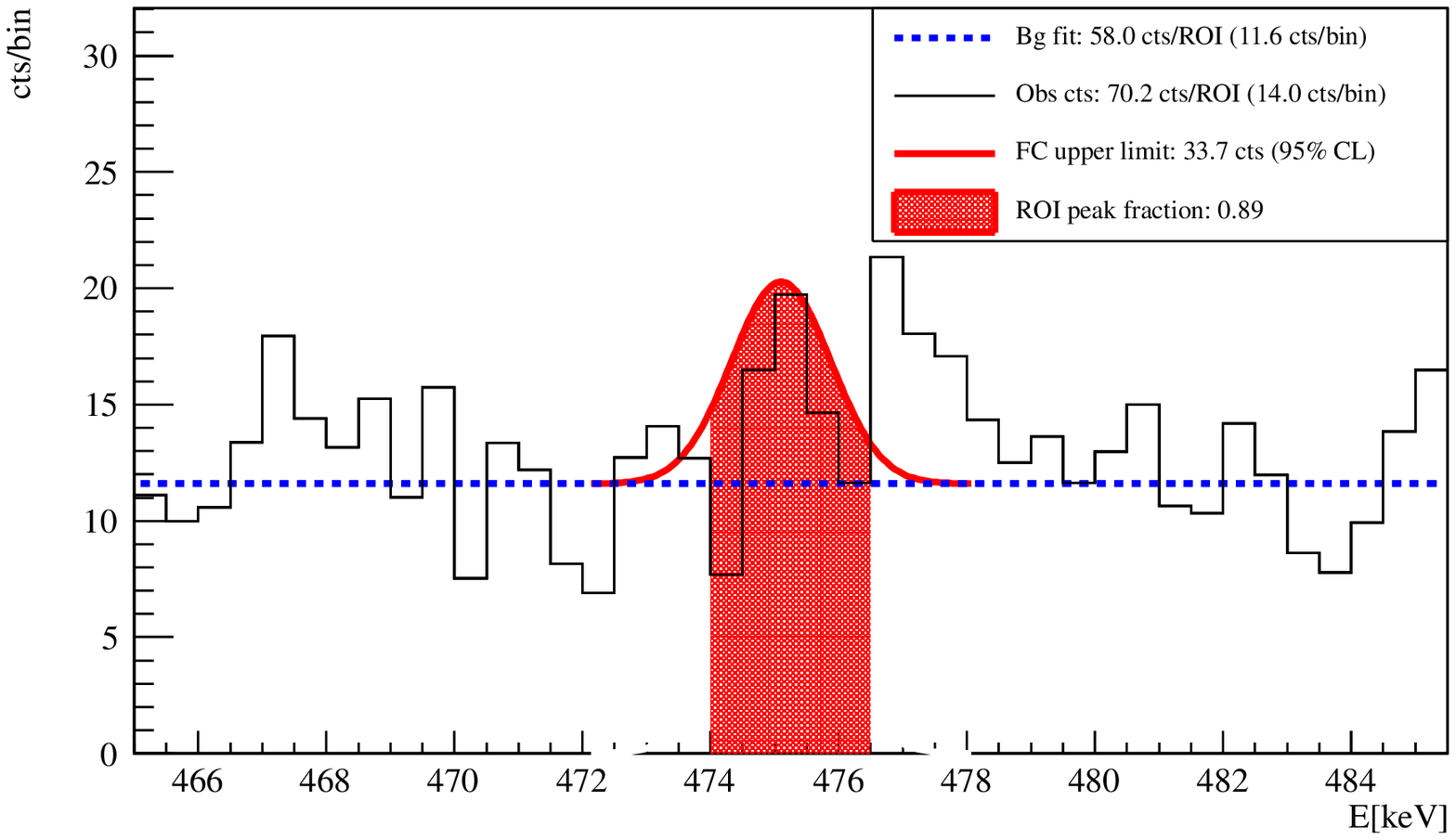}
 \caption{ \label{fig:475keV} The peak region around the \unit[475.10]{keV} \gline\ from \nuc{Pd}{102} $2_1^+$.}
 \end{figure}

\begin{figure}[h]
 \includegraphics[width=0.9\textwidth]{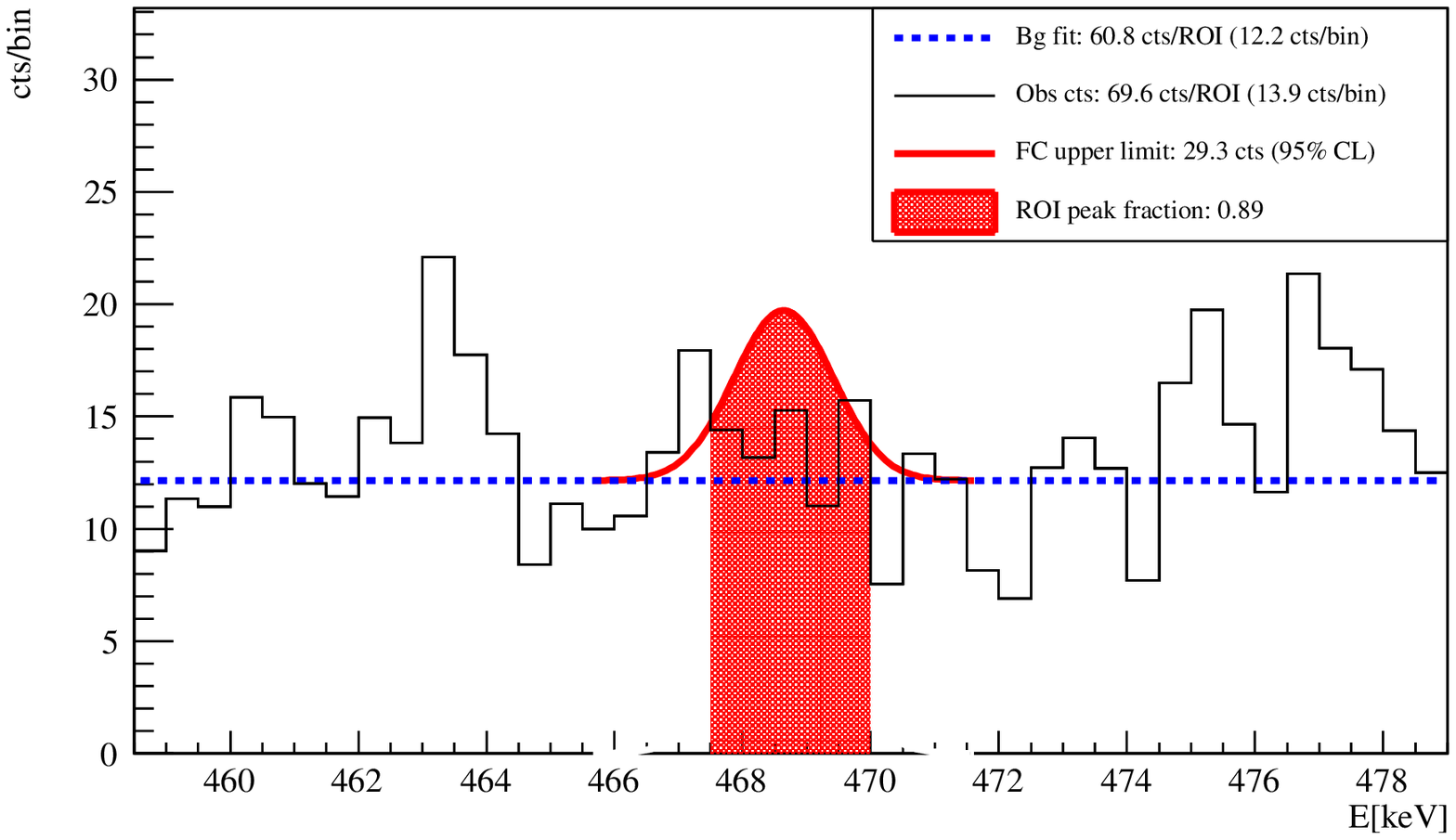}
 \caption{ \label{fig:468keV} The peak region around the \unit[468.64]{keV} \gline\ from \nuc{Pd}{102} $0_1^+$.}
 \end{figure}

All \glines\ participating in a $\gamma$-cascade are investigated and a half-life was calculated for each. Intrinsically, the calculation of limits is influenced by statistical fluctuations in the experimental spectrum, hence the largest calculated limit for an excited state transition is quoted as the half-life limit of this transition. A summary of all investigated \glines\ can be found in \tab \ref{tab:resultsDetailed}: Quoted are the emission probability, the detection efficiency including summation effects, the upper count limit in the spectrum and the calculated half-life. The selected half-life for each transition are summarized in \tab \ref{tab:resultsSummarized}. The peak regions for the \glines\ originating from the favored $0^+_1$ transitions are shown in \fig \ref{fig:657keV} and \ref{fig:815keV} for \nuc{Pd}{110} and in \fig \ref{fig:475keV} and \ref{fig:468keV} for \nuc{Pd}{102}.

\begin{table}[h]
\caption{\label{tab:resultsDetailed} Experimental results for each decay mode and \gline. The columns from left to right denote the decay mode, the \gline\ energy, the emission probability in \% and the detection efficiency in \%. The last two columns show the upper signal count limit and the deduced lower half-life limit at \unit[95]{\%} CL.\\
}
\begin{ruledtabular}
\scriptsize
\begin{tabular}{lrrrrr}
 Decay Mode & \gline\ Energy [keV]  & Emission Probability & Detection Efficiency & Signal Count Limit & \THL\ Limit [yr] \\
\hline
\nuc{Pd}{110} $2^+_1$ \unit[657.76]{keV} & \unit[657.76]{keV} & \unit[100]{\%} &  \unit[4.70]{\%} & 12.4 & \baseT{1.72}{20} \\
\hline
\nuc{Pd}{110} $0^+_1$ \unit[1473.12]{keV} & \unit[815.33]{keV} & \unit[100]{\%} &  \unit[3.84]{\%} & 8.4 &\baseT{1.98}{20} \\
					  & \unit[657.76]{keV} & \unit[100]{\%} &  \unit[3.94]{\%} & 12.4 &\baseT{1.44}{20} \\
\hline
\nuc{Pd}{110} $2^+_2$ \unit[1475.80]{keV} & \unit[1475.80]{keV} & \unit[35.25]{\%} &  \unit[1.32]{\%} & 11.5 & \baseT{5.17}{19} \\
					  & \unit[818.02]{keV} & \unit[64.75]{\%} &  \unit[2.40]{\%} & 16.3 &\baseT{6.67}{19} \\
					  & \unit[657.76]{keV} & \unit[64.75]{\%} &  \unit[2.53]{\%} & 12.4 &\baseT{9.26}{19} \\
\hline
\nuc{Pd}{110} $0^+_2$ \unit[1731.33]{keV} & \unit[1073.7]{keV} & \unit[86.73]{\%} &  \unit[1.89]{\%} & 10.1 & \baseT{8.50}{19} \\
					  & \unit[657.76]{keV}\footnotemark[1] & \unit[95.32]{\%} &  \unit[3.78]{\%} & 12.4 & \baseT{1.38}{20} \\
					  & \unit[255.49]{keV} & \unit[13.27]{\%} &  \unit[0.36]{\%} & 25.3 &\baseT{6.46}{18} \\
					  & \unit[1475.80]{keV} & \unit[4.68]{\%} &  \unit[0.12]{\%} & 11.5 &\baseT{4.87}{18} \\
					  & \unit[818.02]{keV} & \unit[8.59]{\%} &  \unit[0.24]{\%} & 16.3 &\baseT{6.63}{18} \\
\hline
\nuc{Pd}{110} $2^+_3$ \unit[1783.48]{keV} & \unit[1783.48]{keV} & \unit[21.57]{\%} &  \unit[0.88]{\%} & 6.2 & \baseT{6.45}{19} \\
					  & \unit[1125.71]{keV} & \unit[78.43]{\%} &  \unit[2.48]{\%} & 12.0 &\baseT{9.41}{19} \\
					  & \unit[657.76]{keV} & \unit[78.43]{\%} &  \unit[2.99]{\%} & 12.4 &\baseT{1.09}{20} \\
\hline
\hline
\nuc{Pd}{102} $2^+_1$ \unit[475.10]{keV} & \unit[475.10]{keV} & \unit[100]{\%} &  \unit[5.09]{\%} & 33.7 & \baseT{5.95}{18}\\
\hline
\nuc{Pd}{102} $0^+_1$ \unit[943.69]{keV} & \unit[468.64]{keV} & \unit[100]{\%} &  \unit[4.32]{\%} & 29.3 & \baseT{5.81}{18} \\
					 & \unit[475.10]{keV} & \unit[100]{\%} &  \unit[4.31]{\%} & 33.7 &\baseT{5.04}{18} \\
\hline
\nuc{Pd}{102} $2^+_2$ \unit[1103.05]{keV} & \unit[1103.05]{keV} & \unit[37.11]{\%} &  \unit[1.60]{\%} & 13.5 & \baseT{4.66}{18} \\
					  & \unit[627.94]{keV} & \unit[62.90]{\%} &  \unit[2.54]{\%} & 11.7 &\baseT{8.55}{18} \\
					  & \unit[475.10]{keV} & \unit[62.90]{\%} &  \unit[2.67]{\%} & 33.7 &\baseT{3.13}{18} \\
\hline
\end{tabular}
\end{ruledtabular}
\footnotetext[1]{This \gline\ is part of two sub-branches starting from the same excited state.}
\end{table}

\begin{table}[h]
\caption{\label{tab:resultsSummarized} Summary of measured half-life limits for all \nuc{Pd}{110} and \nuc{Pd}{102} double beta decay excited state transitions. \\
}
\begin{ruledtabular}
\begin{tabular}{lr}
 Decay Mode  & \THL\ Limit [yr] (\unit[95]{\%})\\
\hline
\nuc{Pd}{110} $2^+_1$  \unit[657.76]{keV} & \baseT{1.72}{20} \\
\nuc{Pd}{110} $0^+_1$  \unit[1473.12]{keV} & \baseT{1.98}{20} \\
\nuc{Pd}{110} $2^+_2$  \unit[1475.80]{keV} & \baseT{9.26}{19} \\
\nuc{Pd}{110} $0^+_2$  \unit[1731.33]{keV} & \baseT{1.38}{20} \\
\nuc{Pd}{110} $2^+_3$  \unit[1783.48]{keV} & \baseT{1.09}{20} \\
\hline
\nuc{Pd}{102} $2^+_1$  \unit[475.10]{keV} & \baseT{5.95}{18}\\
\nuc{Pd}{102} $0^+_1$  \unit[943.69]{keV} & \baseT{5.81}{18} \\
\nuc{Pd}{102} $2^+_2$  \unit[1103.05]{keV} & \baseT{8.55}{18} \\
\end{tabular}
\end{ruledtabular}
\end{table}

\newpage

\section*{Conclusion}

A palladium sample has been investigated for double beta decay transitions into excited states in the low background laboratory HADES. Lower half-life limits could be improved for the $0^+_1$ and $2^+_1$ transitions in \nuc{Pd}{110} and \nuc{Pd}{102} and first limits are established for all possible higher energetic excited state transitions. The best limit could be set for the \nuc{Pd}{110} $0^+_1$ transition with a half-life larger than \unit[\baseT{1.98}{20}]{yr}. The largest improvement compared to previous results is archived for the \nuc{Pd}{110} $2^+_1$ transition with \unit[\baseT{1.72}{20}]{yr} which is an improvement by a factor of 3.9. For the \nuc{Pd}{102} system, the improvements are smaller due to upward fluctuations of the background in the peak region of the \unit[475.10]{keV} and \unit[468.64]{keV} \glines.\\

Possible improvements of the search of double beta decays in palladium is the consideration of X-rays. This requires a different geometric assembly of the palladium plates e.g.\ in a layer around an n-type HPGe detector with thin dead layer. Further improvement can be achieved considering $\gamma-$coincidences using the multi-parameter DAQ system. This would also reduce the muonic background. The intrinsic massive activity of the palladium sample was determined to be \unit[1.7]{mBq/kg} for the \nuc{U}{238} chain and below the detection threshold for the \nuc{Th}{232} chain. Further purifications are not expected to yield significant improvement. However, accumulating storage underground will reduce the general background from cosmic activated radio isotopes in the palladium sample and the measuring system.

\section*{Acknowledgements}
Gerd Marissens is acknowledged for extensive technical support. Furthermore the HADES-staff of EURIDICE is acknowledged.

\bibliographystyle{h-physrev}

\bibliography{./PdHADESPaper_short}{}

\end{document}